\documentclass{article}
\usepackage{arxiv}

\usepackage[utf8]{inputenc} 
\usepackage[T1]{fontenc}    
\usepackage{hyperref}       
\usepackage{url}            
\usepackage{booktabs}       
\usepackage{amsfonts}       
\usepackage{nicefrac}       
\usepackage{microtype}      
\usepackage{lipsum}		
\usepackage{graphicx}
\usepackage{subcaption}
\usepackage{comment}
\usepackage{rotating}
\usepackage{tabularx}
\usepackage{makecell}
\usepackage{mathtools}
\usepackage{adjustbox}

\newcommand{\quotes}[1]{``#1''}

\title{A Systematic Review on Context-Aware Recommender Systems using Deep Learning and Embeddings}

\author{
    Igor André Pegoraro Santana \\
	Department of Informactics (DIN)\\
	State University of Maringá\\
    Maringá, Paraná, Brazil \\
	\texttt{pg400816@uem.br} 
	\And
	 Marcos Aurelio Domingues \\
	Department of Informactics (DIN)\\
	State University of Maringá\\
    Maringá, Paraná, Brazil \\
	\texttt{madomingues@uem.br} 
}

\date{}

\hypersetup{
pdftitle={A Systematic Review on Context-Aware Recommender Systems using Deep Learning and Embeddings},
pdfsubject={cs.LG},
pdfauthor={Igor André Pegoraro Santana, Marcos Aurelio Domingues},
pdfkeywords={Systematic review, Recommender systems, Context-aware recommender systems, Embeddings, Deep learning},
}

\begin{document}
\maketitle

\begin{abstract}
Recommender Systems are tools that improve how users find relevant information in web systems, so they do not face too much information. In order to generate better recommendations, the context of information should be used in the recommendation process. Context-Aware Recommender Systems were created, accomplishing state-of-the-art results and improving traditional recommender systems. There are many approaches to build recommender systems, and two of the most prominent advances in area have been the use of Embeddings to represent the data in the recommender system, and the use of Deep Learning architectures to generate the recommendations to the user. A systematic review adopts a formal and systematic method to perform a bibliographic review, and it is used to identify and evaluate all the research in certain area of study, by analyzing the relevant research published. A systematic review was conducted to understand how the Deep Learning and Embeddings techniques are being applied to improve Context-Aware Recommender Systems. We summarized the architectures that are used to create those and the domains that they are used.
\end{abstract}


\section{Introduction}

\par The use of Recommender Systems (RS) as a technique to filter information that matches the preferences from a user has increased as the amount of information available to the user grow as well. Traditional RS use only information about users and items to generate the recommendations, and they do not take into consideration any contextual information, such as time and place that a user has rated an item \cite{Adomavicius2011}. Context-Aware Recommender Systems (CARS) are algorithms that use information beyond users and items to generate those recommendations, and have been developed by academic researchers and applied in a variety of different applications settings \cite{Adomavicius2011}. 
\par In the past few years due to the decrease in computational cost, Deep Learning (DL) has been applied in many areas of research, due to its capability in solving many complex tasks \cite{Zhang2017}. In \cite{Lecun2015}, DL shows its strength because it \quotes{discovers intricate structure in large data sets by using the backpropagation algorithm to indicate how a machine should change its internal parameters that are used to compute the representation in each layer from the representation in the previous layer}. These representations of the input may have features that describe the input value. As well as DL, Representation Learning (RL) techniques, in special Word Embeddings, has been getting attention due to its effectiveness in storing valuable syntactic and semantic information
 \cite{Camacho-Collados2018}. Embeddings are a set of techniques that are used to reduce the dimensionality and generate an alternate representation of the input. Both areas are being used to build RS that tries to solve some of the well-known problems (i.e. cold-start and data sparsity) while obtaining state-of-the-art results. Although there are surveys that review DL based RS \cite{Zhang2017, Liu2017, Tilahun2017}, there are a lack of systematic reviews that reviews CARS that have been built using DL techniques and  that use information that is obtained by Embeddings models. The goal of this systematic review is to review literature that proposes DL and Embeddings models to advance the state-of-the-art of CARS in different domains. The remaining of this paper is organized as follows: The section Background Concepts contain concepts about Embeddings, Deep Learning, CARS and Systematic Reviews that are needed to understand how the systematic review was applied in this work. Then, the protocol used in the review is detailed in the Systematic Review section, followed by the section Results, which discloses the results obtained in the review. Finally, we present some conclusions.

\section{Background Concepts} 

\par In this section, we present some key concepts that are related to our systematic review, as well as the concepts that are related to the systematic review itself.

\subsection{Deep Learning}
\par DL techniques are making major advances in problems that machine learning did not have good results. They are good at discovering intrinsic structures in data that are high-dimensional and applicable to many domains \cite{Lecun2015}. A neural network is said to be \textit{deep} if it has layers between the input and output layer. Although we call the technique DL, there are different models to be applied to each individual problem. 
\par A model that achieve state-of-the-art results \cite{Krizhevsky12} is the Convolutional Neural Network (CNN). It is designed to process multiple arrays as input, with arrays ranging from 1 dimension (signals and sequences) to 3 dimensions (video or volumetric images). In \cite{Lecun2015}, the architecture for a CNN is structured as a series of stages, being the first stages with two types of layers: convolutional and pooling layers. The role of the convolutional layer is to detect local conjunctions of features from the previous layer, and the pooling layer is to merge semantically similar features into one. 
\par A DL model that is usually applied to RS is the Recurrent Neural Network (RNN), and they are specialized in dealing with tasks that involve sequential inputs. They process an input sequence one element at a time, maintaining information about the information that was already processed in the hidden units. Hence, they use information from past iterations to predict the output \cite{Lecun2015}.

\subsection{Embeddings}

\par Vector representations of discrete items are being used in distinct areas of studies, mainly because they help solve the problem of the curse of dimensionality \cite{Bengio2003}, which happens when the number of data available grows, increasing the sparsity of the data set, requiring more storage and processing time. The curse of dimensionality can be seen in different areas, as Natural Language Processing \cite{Bengio2003}, RS \cite{Amatriain2015} and Data Mining \cite{Verleysen2005}.

\par One of the first works that tackled the problem of the curse of dimensionality by generating distributed representations is \cite{Bengio2003}, which used a FeedForward Neural Network with a linear projection layer to jointly learn the representations of words and a statistical language model. A common model that is used to generate the representations are the Autoencoders networks. Autoencoders are networks with $n$ units in the input and the output layer, and $p$ units in the hidden layer, where $p < n$ and the main goal is to produce the same result in the output that was used as input. With one linear hidden layer and an error criteria (e.g. mean squared error) to train the network, the $p$ hidden units learn how to project the input value to the output layer. The weights of the $p$ hidden units can be used as representation of the input with less dimensions \cite{Bourlard1988}.

\par Recent advances in Embeddings were presented by \cite{mikolov}, with a model called Word2Vec. It was seen that beyond the distributed representations, the model proposed was able to generate representations that are semantically valuable. If algebraic operations are performed on the word vectors, it results in a vector representation closest to what is desirable. As an example, the following operation $ \text{vector}(\text{\quotes{King}}) - \text{vector}(\text{\quotes{man}}) + \text{vector}(\text {\quotes{woman}})$ results in the vector representation of the word $\text{vector}(\text{\quotes{Queen}})$.

\subsection{Context-Aware Recommender Systems}

\par Traditional Recommender Systems only use information about the user and the items which were visited by the user in order to recommend new items. Thus, they do not take into account contextual information to generate those recommendations, such as time, place and the sequence of items visited by the user \cite{Adomavicius2011}. The use of contextual information is important because a user may prefer different kinds of items depending on which time of the day he/she is, or with whom he/she is with. In \cite{Adomavicius2005}, it is said that \quotes{accurate prediction of consumer preferences undoubtedly depends upon the degree to which the recommender system has incorporated the relevant contextual information into a recommendation method.}
\par There are multiple ways to incorporate the contextual information into the recommendations process that can be categorized based on two main aspects \cite{Adomavicius2011}:
\begin{enumerate}
    \item What a recommender system may know about the contextual factors;
    \item How contextual factors change over time.
\end{enumerate}
\par The first aspect presumes that a recommender system has information about the contextual factors, and the recommenders are classified depending on how much they know about the factors. For the first aspect, there are three categories:
\begin{itemize}
    \item \textit{Fully observable}: All the contextual factors, as well as their structure and values that are relevant to the application are known explicitly by the recommender system;
    \item \textit{Partially observable}: Only some of the information are known explicitly. As an example, the recommender system may know the contextual factors, but not know their structure;
    \item \textit{Unobservable}: The recommender system has no information about the contextual factor, and it makes the recommendations by using latent knowledge about the data.
\end{itemize}

The second aspect is whether and how the structure of the contextual factors change over time. There are two categories that the recommenders can be classified depending on how the factors change:
\begin{itemize}
    \item \textit{Static}: The relevant contextual factors do not change over time;
    \item \textit{Dynamic}: The contextual factors can change over time. As an example, the recommender system may realize that some factors are not relevant to generate recommendations over time, and not use it anymore.
\end{itemize}

\par These categories are used to describe how the CARS interacts with the contextual information, and how it will obtain those information. 

\subsection{Systematic Review}

\par A systematic literature review (often called systematic review), as described in \cite{Kit04}, is adopted to identify, evaluate and to interpret the relevant research available in a particular research area. A systematic review must follow pre-defined phases to review the literature, with the intent of being reproducible. It has three defined phases: planning the review, conducting the review and reporting the results obtained. In the first phase, a research protocol is defined that contains the questions that the systematic review wants to answer as well as the research strategies. The second phase is responsible for applying the research protocol and extracting the results from it. Finally, the results are obtained and it is defined how they will be reported. Systematic reviews only review \textit{primary works}, that are references that contribute or modify the state-of-the art of a research area. Surveys and systematic reviews are called \textit{secondary works}, and they are not reviewed in the systematic review. 

\subsubsection{Planning}

\par The planning phase of a systematic review is responsible for the identification of the need of the review and the development of the review protocol. In \cite{Kit04}, the tasks in the planning phase are described as:
\begin{itemize}
    \item Identification of the need of the review: the need for a systematic review comes from the requirement to summarize information in a thorough and unbiased manner;
    \item Formulation of the research question: the most important part of a systematic review, because it guides the entire process. A critical issue in any systematic review is to ask the right question, based on the need;
    \item Development of the protocol: the critical element of any systematic review, because the review must be reproducible. So, the protocol must have well-defined information.
\end{itemize}

\subsubsection{Conduction}

\par After the protocol is completed, the actual review starts. The tasks that comprehend the conduction phase are described in \cite{Kit04} as the following:
\begin{itemize}
    \item Searching the references: in this task the research question are applied in data sources that contains references, by using search queries;
    \item Research selection: once the relevant references has been gathered, they need to be assessed for their relevance. Works that were found in the search must be selected based on inclusion/exclusion criteria;
    \item Quality assessment (optional): in addition to the inclusion/exclusion criteria, the quality of the reference can be assessed;
    \item Data extraction: the objective of this task is to obtain useful information from the obtained work. These extracted information must answer the research questions defined previously;
    \item Data synthesis: this task involves summarizing the results of the data extraction task.
\end{itemize}

\subsubsection{Report}

\par After all the references are obtained from the conduction phase, the results must be communicated. As seen in \cite{Kit04}, there are at least two formats in which the review can be reported:

\begin{enumerate}
    \item Technical report or in a section of a PhD thesis;
    \item Journal or conference paper.
\end{enumerate}

\par Moreover, the results and the information obtained by the systematic review must be understandable to the reader, with insights about the references and graphs so the reader can visualize what is in the references.

\section{Systematic Review} 

\par The application of the systematic review in this work has the objective of studying how DL and Embeddings are being applied in state-of-the-art CARS algorithms. Every systematic review has questions that the obtained works must answer. In this work, our research questions are the following:
\begin{itemize}
    \item What models are being used to generate Embeddings for CARS?
    \item What models are being used to recommend items with DL based CARS?
    \item In which domains are those models being applied?
\end{itemize}

\par After defining the research questions that the systematic review must answer, we begin the planning phase of the systematic review. The main intent of this work, is to summarize the results published in the CARS area that use DL and Embeddings models. There are reviews that summarize works done in traditional RS \cite{Zhang2017, Liu2017, Tilahun2017}, but not reviews on CARS. To the best of our knowledge, there are no reviews done in the area of Embeddings and DL on CARS.

\par As the research questions are created and the need for the systematic review is justified, the development of the protocol can start. The protocol must have information about how the references are going to be searched, and the first step is to create research queries that are used to search the data sources for the references. To create those questions, we enumerated terms that are necessary to be found in the references: \textit{deep learning}, \textit{context-aware recommender systems} and \textit{embeddings}. The data sources used to perform the reference search have slightly differences in their syntax, but a generic query can be seen as the following: 

\textit{(\quotes{deep learning} AND \quotes{context-aware recommender systems}) OR  (\quotes{embeddings} AND \quotes{context-aware recommender systems})}

\par The objective of this query is to retrieve references that contains the words \textit{context-aware recommender systems} together with \textit{deep learning} or \textit{embeddings}. The search query was applied in different data sources that contains references in the computer science area. The data sources used in this work are the following:
\begin{itemize}
    \item ACM Digital Library {\url{https://dl.acm.org}}
    \item IEEE Xplore Digital Library {\url{https://ieeexplore.ieee.org}}
    \item ScienceDirect {\url{https://www.sciencedirect.com}}
    \item Springer Link {\url{https://link.springer.com}}
\end{itemize}

\par Then, the next step of the protocol is to define the selection criteria that is used in the conduction phase. These criteria are applied to the references retrieved from the data sources in order to filter the references that did not match our goal. Only the references that did not match the following exclusion criteria are maintained:
\begin{itemize}
    \item \textbf{C1:} References that have restricted access or that are not written in English;
    \item \textbf{C2:} References that have less than 4 pages, posters, presentations, and tutorials;
    \item \textbf{C3:} References that are secondary works (reviews and surveys);
    \item \textbf{C4:} References that do not propose an approach that aims to improve a context-aware recommender system by using DL or Embeddings.
\end{itemize}

\par To finish the planning protocol, it is important to define a set of information that is going to be extracted from each reference that was retrieved, as follows:
\begin{itemize}
    \item Basic information about the reference, e.g. title, authors, year and where it was published;
    \item Which dataset was used to perform the experiments;
    \item If it was used DL or Embeddings;
    \item Architecture used on DL and Embedding models;
    \item Domains that the recommendations are made;
    \item Contextual information used in the recommendation process;
    \item How the contextual information was used (contextual approach);
    \item How the results were evaluated (evaluation protocol, metrics, etc);
    \item Which algorithms were used as baseline.
\end{itemize}

\par Once the research protocol was defined, the conduction phase of the systematic review was started. The first step of the conduction phase is to apply the search query to the data sources that contain the references. The results of the search in the data sources can be seen in Table~\ref{table:datasets}. As we can see in the results, ACM Digital Library has the most amount of works found in every data source. This can be explained due to the most famous conference in the RS area, RecSys \footnote{https://recsys.acm.org} is hosted by ACM, therefore it is expected to have more results. 

\begin{table}[!ht]
    
    \begin{adjustbox}{width=.5\textwidth,center}
        \begin{tabular}{ll}
            \hline
            Data source & Number of works \\ \hline
            ACM Digital Library & 193 \\
            IEEE Xplore Digital Library & 30 \\
            ScienceDirect & 29 \\
            Springer Link & 20 \\ \hline
            Total & 272 \\ \hline
        \end{tabular}
    \end{adjustbox}
    \caption{Number of works found in each data source.}
    \label{table:datasets}
\end{table}

\par To organize the references obtained from the data sources, an application that manages references called Mendeley\footnote{https://www.mendeley.com} was used. The application is useful for removing duplicates, organizing and downloading the references as a PDF file. 

\par After all the references were placed in Mendeley, we started to select the references that did not match with the exclusion criteria. Before that process, the duplicate references were removed, as there was an intersection of works between the data sources. We have removed 41 duplicated references, with a new total of 231 references. Table~\ref{table:criteria} summarizes the number of references obtained after applying the exclusion criteria defined in the planning phase. 

\begin{table}[!ht]
    \begin{adjustbox}{width=.5\textwidth,center}
        \begin{tabular}{lr}
        \hline
        \textbf{Exclusion criteria} & \textbf{References} \\ \hline
        Total of References         & 272                           \\
        After elimination of duplicates   & 231                           \\
        After applying the exclusion criteria C1           & 227                           \\
        After applying the exclusion criteria C2           & 184                           \\
        After applying the exclusion criteria C3           & 156                           \\
        After applying the exclusion criteria C4           & 16                            \\ \hline
        \end{tabular}
    \end{adjustbox}
    \caption{Number of references after each step of the selection.}
    \label{table:criteria}
\end{table}
\par As we can see in Table~\ref{table:criteria}, most references were removed due to they do not have any proposal of a Context-Aware Recommender System. This check was done by searching the abstract of the references to understand the proposal made by the authors. In some cases, the abstract did not have enough information, so the full text was read to confirm the exclusion.

\section{Results} 
\label{sec:res}
\par In this section, we focus on the 16 references that were obtained as the results of the application of our systematic review. Prior to the research questions, we will discuss some insights obtained by the references.

\par
\begin{figure}[!ht]
    \centering
    \includegraphics[scale=.5]{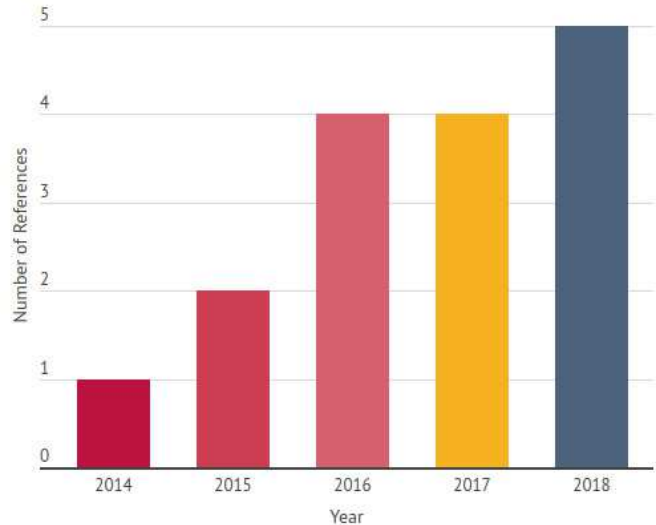}
    \caption{Number of references per year.}
    \label{fig:years}
\end{figure}

 The references tallied by our systematic review are all recent, with the first work \cite{Tang2014} being published in the year of 2014, which is explained by the fact that Embeddings were only proposed in 2013 by \cite{mikolov}. In Figure~\ref{fig:years}, we can see the growth  in the usage of DL and Embeddings techniques to improve CARS. 

\par The techniques that are being used in the publications are fairly new, and because of that, we can see an increase growth of publications with CARS. The year of 2018 has reached the peak in the number of publications so far, which show us that studies in the area tends to grow even more. 

\par
\begin{figure}[!ht]
    \centering
    \includegraphics[scale=.5]{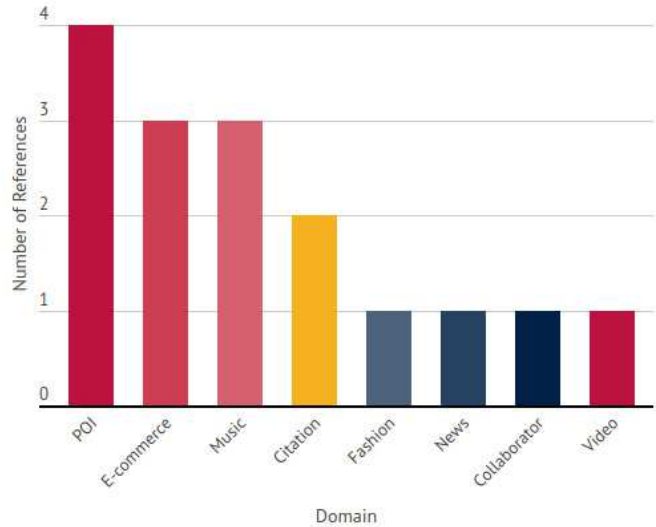}
    \caption{Reference's domains.}
    \label{fig:domains}
\end{figure}

 In \cite{Ricci2015}, RS are defined as \quotes{tools or techniques that are capable of providing items that the user may like}. Item is a generic definition of what can be recommended by the RS, and it varies according to the domain that the RS is used. Although he references found are in different domains of application, they are mainly focused on dealing with sessions. Sessions are a sequence of interactions that a user have in a web application, and CARS that uses session mostly tries to obtain implicit contextual information about the user behavior. However, most of the references do not exploit the session by itself, but the combination of session and another domain (e.g. music, video) in order to make the recommendations. In Figure~\ref{fig:domains} we can see the domains in which the recommendations are made, with the Point of Interest (POI) domain being the one that is most explored, in which the recommendation item is a place or a location that the user may find useful or interesting. 

\par The main focus of research from 9 out of 16 references obtained was to deal with sessions paired with different domains. The E-commerce was the most explored domain amongst the other domains, with 3 references \cite{Tan2016a, Smirnova2017, Li2018}, followed by Music \cite{Wang2016, Wang2018}, POI \cite{Zhao2017}, Fashion \cite{Jannach2017}, News \cite{Zhang2018} and Video \cite{Beutel2018}.

\par
\begin{figure}[!ht]
    \centering
    \includegraphics[scale=.5]{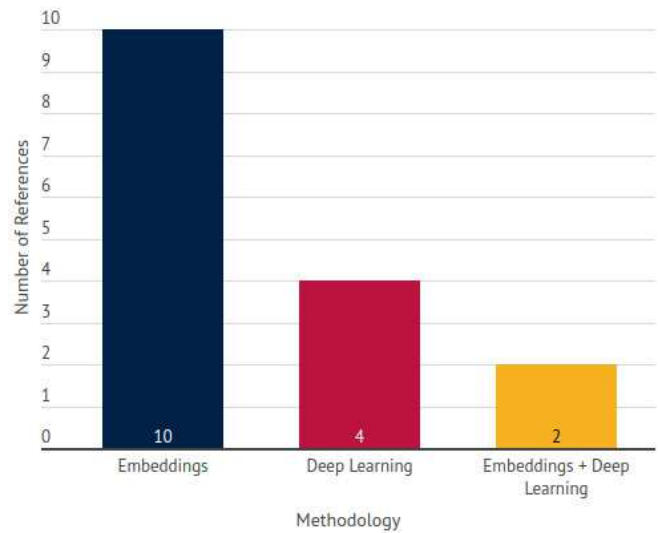}
    \caption{Reference's techniques.}
    \label{fig:methods}
\end{figure}

 From the references obtained, we can see that Embeddings is more popular than DL for building CARS, as we could gather much more references. As shown in Figure~\ref{fig:methods}, in 10 out of 16 references used Embeddings to build its CARS, 4 used DL techniques and 2 references explored the combination of both techniques. 

\par Although there are more references that use Embeddings than Deep Learning, an interest in the combination of both techniques are rising as they become more popular. \cite{Tan2016a} and \cite{Li2018} used Embeddings techniques as input to the DL architecture, resulting in better results that pure DL architectures.
\par In the following subsections, we will discuss the Embeddings and DL models that were used in the references obtained by the systematic review. A brief discussion of the model and how it was applied in each reference is presented. In Table~\ref{quadro:summary}, we present a summary of the references that were obtained by the systematic review, which contains the datasets that were used by the references, the domain that was explored and what contextual information where used in the CARS.

\begin{table}[!ht]

\begin{adjustbox}{width=\textwidth,center}
\begin{tabular}{|l|l|l|l|l|}
\hline
\textbf{Reference}                    & \textbf{Dataset}         & \textbf{Type of model}       & \textbf{Domain}        & \textbf{Contextual Information}                 \\ \hline
\cite{Tang2014}     & Chinese Journal of Computers   & Embeddings                 & Citation                & Words around citation             \\ \hline
\cite{Deng2015}     & SinaWeibo                      & Embeddings                 & Music                   & Emotion                           \\ \hline
\cite{Unger2015}    & POIs em Bersebá, Israel        & Embeddings                 & POI                     & Mobile sensor data                \\ \hline
\cite{Tan2016}      & Library of Quotes              & Embeddings                 & Citation                & Words around citation             \\ \hline
\cite{Tan2016a}     & RecSys 2015 Challenge          & \makecell[bl]{ Embeddings and \\ Deep Learning} & E-commerce              & Click sequence                    \\ \hline
\cite{Xie2016}      & Foursquare e Gowalla           & Embeddings                 & POI                     & \makecell[bl]{Geo-referenced data, Time \\ and Semantic Information}    \\ \hline
\cite{Wang2016}     & Xiami Music                    & Embeddings                 & Music                   & \makecell[bl]{Sequence, Time and Location}                        \\ \hline
\cite{Yang2017}     & Gowalla e Yelp                 & Embeddings                 & POI                     & \makecell[bl]{Geo-referenced data and \\ Social Network Information}    \\ \hline
\cite{Smirnova2017} & YooChose e proprietary data    & Deep Learning              & E-commerce              & Time and Type of interaction                          \\ \hline
\cite{Zhao2017}     & Foursquare e Gowalla           & Embeddings                 & POI                     & Sequence of POIs                                      \\ \hline
\cite{Jannach2017}  & Zalando                        & Deep Learning              & Fashion                 & Features and Click sequence                           \\ \hline
\cite{Zhang2018}    & Adressa 16G e Last.fm          & Deep Learning              & News                    & Time                                                  \\ \hline
\cite{Liu2018}      & AMiner                         & Embeddings                 & Collaborator            & Collaboration topics                                  \\ \hline
\cite{Beutel2018}   & Youtube                        & Deep Learning              & Video                   & Time, Device and page                                 \\ \hline
\cite{Li2018}       & Tianchi e JD                   & \makecell[bl]{ Embeddings and \\ Deep Learning} & E-commerce              & Time                                                 \\ \hline
\cite{Wang2018}     & Xiami Music                    & Embeddings                 & Music                   & Time and Device                                       \\ \hline
\end{tabular}
\end{adjustbox}
\caption{Summary of the references obtained by the Systematic Review.}
\label{quadro:summary}
\end{table}

\subsection*{Embeddings models}

\par The models in the references are used to obtain a distributed representation that will be used as the input of the CARS. From the references, it is possible to obtain three main categories of models that are used: Graph-based models, that explore the weights of the edges of a graph, Neural Networks Embeddings models, that obtain the representations through a process of optimization done by a neural network, and Natural Language Processing (NLP) based models, that use techniques that are used to represent texts and documents as vectors. 

\par The only reference that proposed a model to generate embeddings by using graphs is \cite{Xie2016}. They proposed to generate recommendations by using a bipartite graph. A graph $G = (V, E)$ is called bipartite if its vertices $V$ can be partitioned in 2 classes such that every edge has its ends in different classes \cite{Diestel2017}. As bipartite graphs have 2 classes, \cite{Xie2016} built four bipartitite graphs: POI-POI, POI-Region, POI-Time and POI-Word. Those graphs map POI to different contextual information, and the embeddings that are generated represent different types of vertices in those graphs. Thus, they captured sequential effect with the POI-Time graph, the geographical influence with the POI-Region and the semantic effect with the POI-Word by embedding the graphs into a share low dimensional space, and using the relations between the nodes to generate the recommendations. Those effects influence the recommendation, just like in a traditional CARS.

\par
\begin{figure}[!ht]
    \centering
    \begin{subfigure}[b]{.49\textwidth}
        \includegraphics[width=\textwidth]{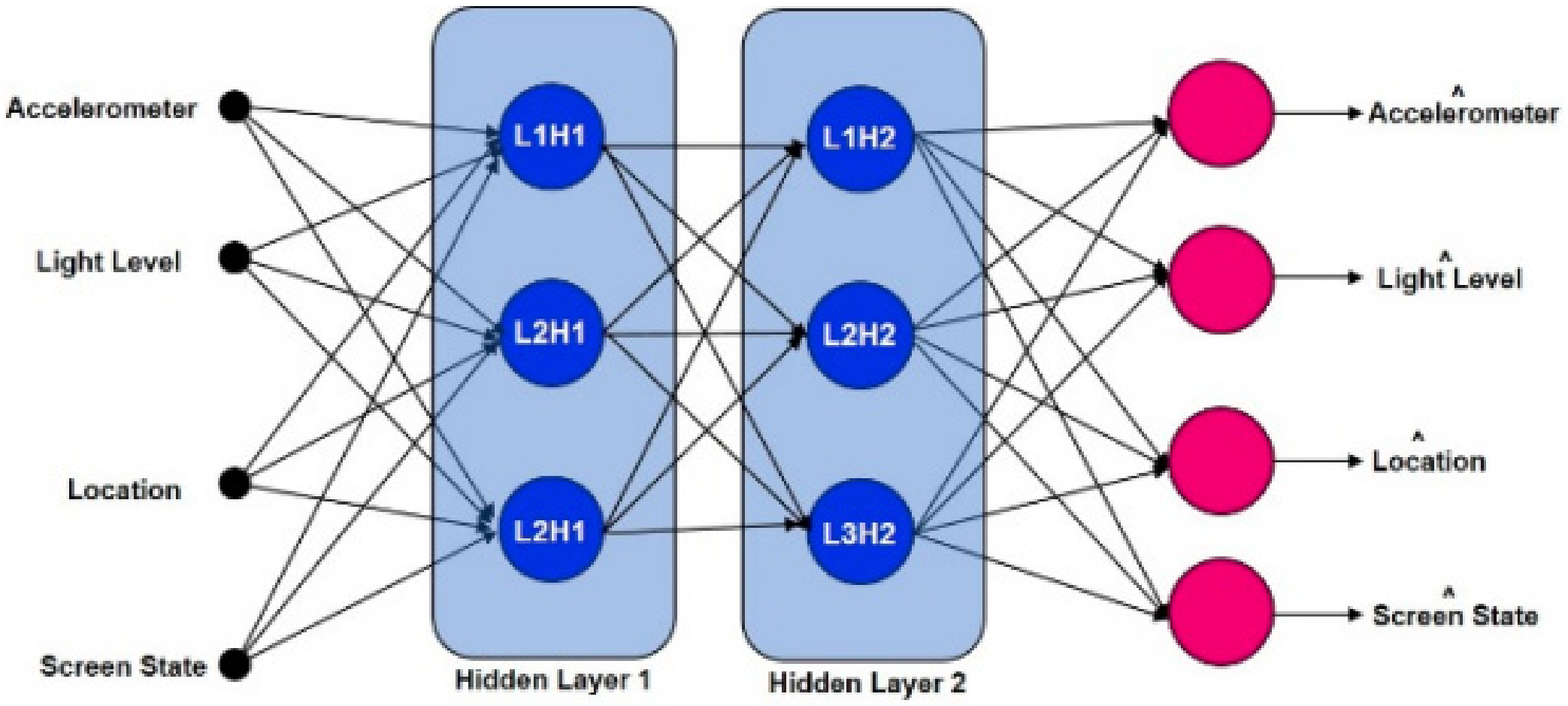}
        \caption{Usage of an Autoencoder  to generate the Embeddings vector\cite{Unger2015}.}
        \label{fig:unger}
    \end{subfigure}
    ~
    \begin{subfigure}[b]{.49\textwidth}
        \includegraphics[width=\textwidth]{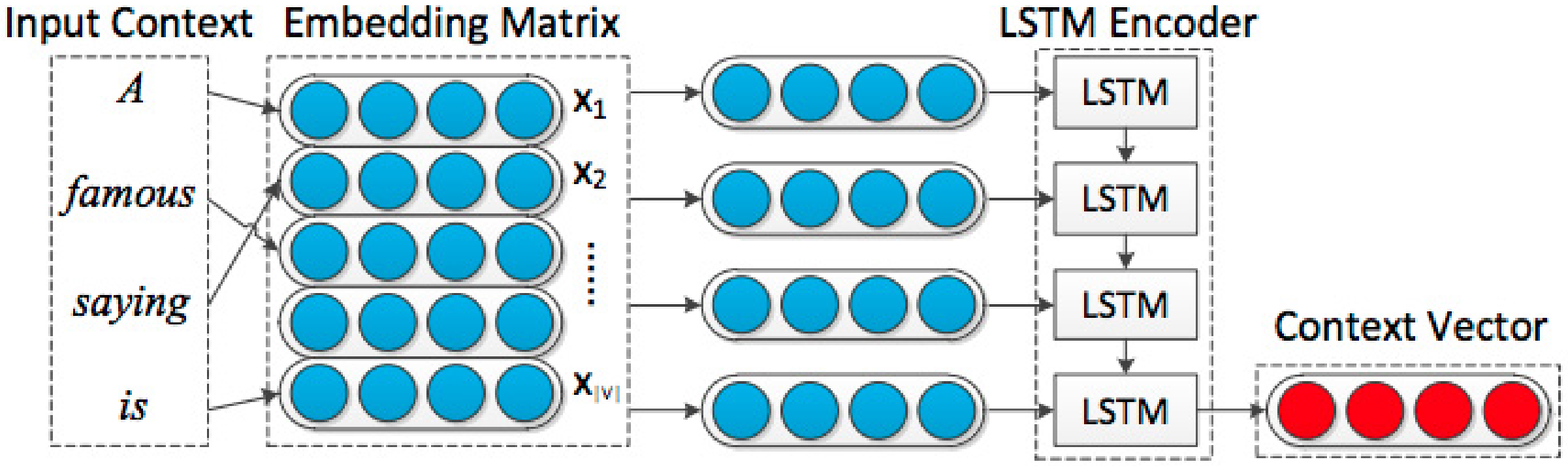}
        \caption{Architecture of the model that converts the context into a vector \cite{Tan2016}.}
        \label{fig:tan}
    \end{subfigure}
    
    ~ 
    \begin{subfigure}[b]{.48\textwidth}
        \includegraphics[width=\textwidth]{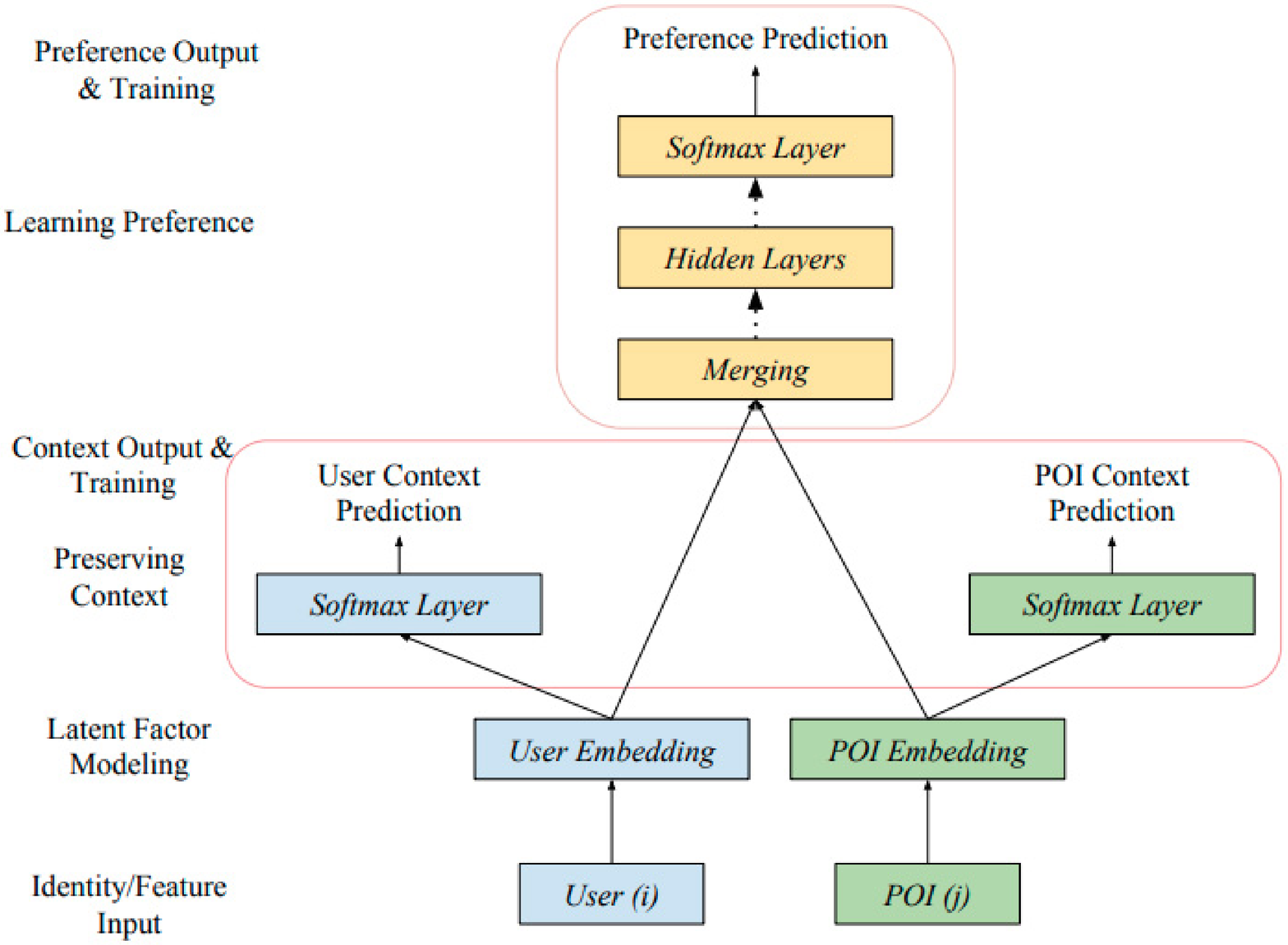}
        \caption{Merging of the embeddings from an user and a POI to generate the recommendations \cite{Yang2017}.}
        \label{fig:yang}
    \end{subfigure}
     ~
    \begin{subfigure}[b]{.48\textwidth}
        \includegraphics[width=\textwidth]{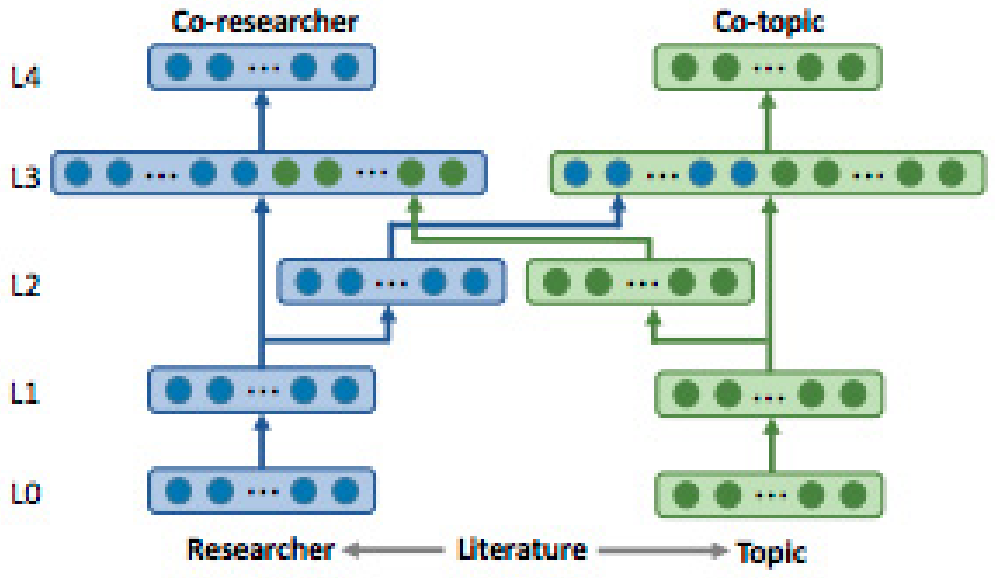}
        \caption{Dual-architecture of the recommender, in which the embedding vector is generated in layer L1 \cite{Liu2018}.}
        \label{fig:liu}
    \end{subfigure}
    \caption{Models that use Embeddings to generate the recommendations.}\label{fig:emb_models}
\end{figure}

 The references that proposed approaches based on Neural Networks used different types of neural networks to generate the distributed representations. \cite{Unger2015} used Autoencoder neural networks, that are unsupervised learning algorithms that applies backpropagation to reduce the error. The main goal is to produce in the output layer the same value given as input, and using the weights in the inner layer to represent the input with fewer dimensions, as it can be seen in Figure~\ref{fig:unger}. 

\begin{figure}[!ht]
    \includegraphics[width=\textwidth]{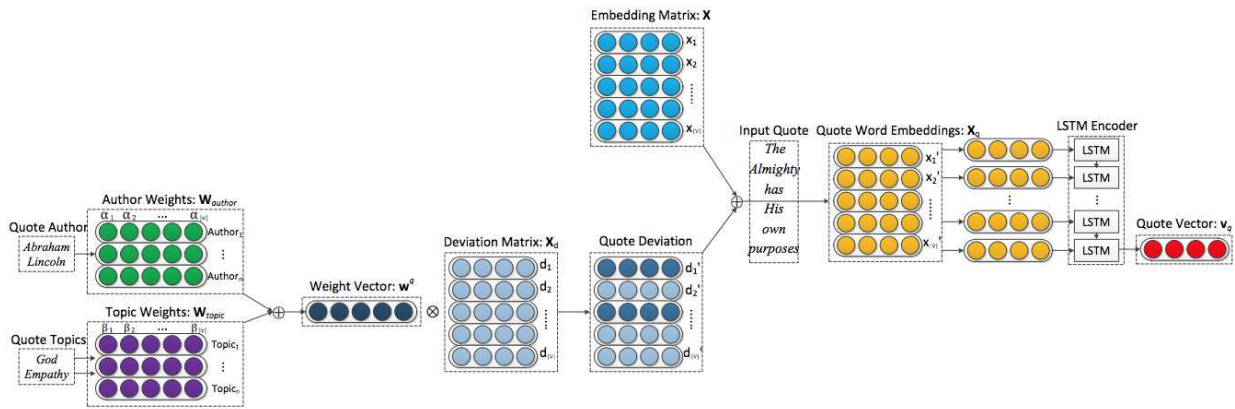}
    \caption{The entire Embeddings architecture, that converts a quote into a vector \cite{Tan2016}.}
    \label{fig:tanquote}
\end{figure}

\par In \cite{Tan2016}, an LSTM encoder was proposed to learn the semantic representations of quotes in the context in which they could be used. It used two different encoders: a context encoder Figure~\ref{fig:tan}, which encodes the sequence of words in a paragraph into a fixed-dimensional representation vector, and a quote encoder Figure~\ref{fig:tanquote}, that maps an quote into a fixed-dimensional representation vector. Quotes as slightly different from paragraphs, as the words used in it might have different meanings.

\par The work developed by \cite{Liu2018} tries to recommend collaborators that are likely to work with a researcher in a given topic. It does so by using a model called Collaborative Entity Embedding, which maps topics and researchers in compact vector spaces. These vector spaces contain the researchers collaboration tendencies and topics underlying semantics. It adopts a dual-structure, as it can be seen in Figure~\ref{fig:liu}, which preserves the mutual-dependency while modeling entities’ (i.e. researchers and topics) co-occurrence relationship.

\par Finally, the work developed by \cite{Yang2017} uses Deep neural networks to model the user preference for a given POI. Although they generate representations to represent the users and the items, the context preference is obtained by a NLP-based model called Skip-gram. In Figure~\ref{fig:yang} we have the overall architecture used to generate the recommendations.

\par The remaining references that use Embeddings are based on NLP models. Most of the works are based on the models proposed by \cite{mikolov}, which maps words into word vectors. However, the references \cite{Tang2014, Deng2015} that were first published regarding the usage of embeddings in CARS used the Bag-of-Words technique to create the vectors. \cite{Tang2014} extends a model proposed by \cite{weston2010}, which uses bags-of-visual terms to represent images. Instead of using visual terms, they obtain terms by the annotations, which are then used in the process of recommending citation in scientific articles. The reference maps the contexts and the citations into the same low dimensional space, with the assumption that if a context cites several papers, the papers should have the similar topic and be mapped into similar vectors.

\par In \cite{Deng2015}, the bag-of-words model is used to map emotions obtained by a classification system into a vector representation, and the length of the vector is based on the number of emotions that can be classified by the system. That way, they use the emotions as context to improve the music recommendation process.

\par
\begin{figure}[!ht]
    \centering
    \includegraphics[scale=.5]{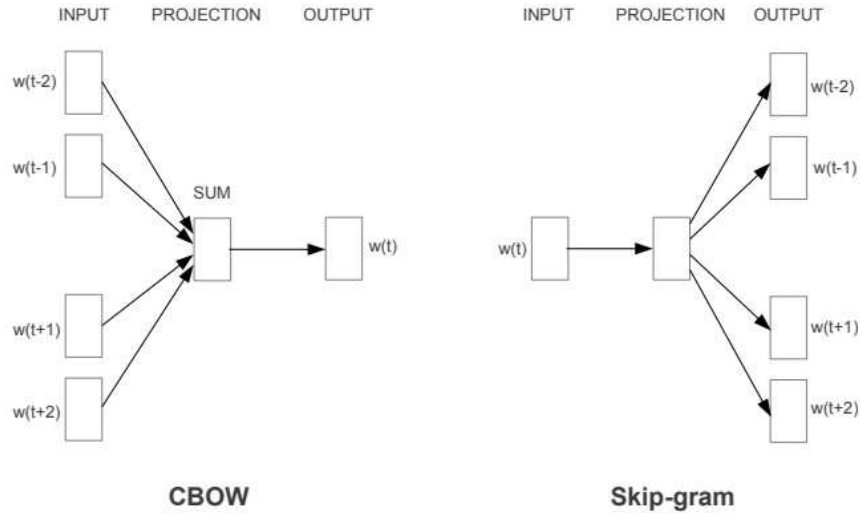}
    \caption{Models proposed by \protect  \cite{mikolov}}
    \label{fig:mikolov}
\end{figure}

 The remaining models are based on the Word Embeddings models proposed by \cite{mikolov}: Continuous Bag-of-Words (CBOW) and Continuous Skip-Gram. In Figure~\ref{fig:mikolov}, we show the architecture used on both models. In the CBOW model, the projection layer is shared for all words, and, because of that, all words are projected into the same position and their vectors are averaged. In this model, the order of the words does not influence the projection. The Continuous Skip-gram model is similar to the CBOW, but instead of predicting the word based on the context that the word is, it tries to predict the context based on the word. Each word is used as input to a log-linear classifier with a continuous projection layer to predict words within a certain range.

\par In \cite{Tan2016a}the authors developed a model similar to the one proposed by \cite{mikolov}, and it was used to generate the embeddings that were used as input and output to an DL model. In this case, embeddings were used to reduce the dimensionality of the data from one-hot encoding to a much smaller vector. That way, the goal of the DL model is to predict the embedding which will represent  item to be recommended.

\par In  \cite{Wang2016}, the authors propose a Music Embedding Model, which is based on the CBOW. The model includes metadata from music and additional context information to CBOW, with the goal of predicting the music based on its local context (the words on the input), metadata and additional context information. This model was used to generate the embeddings for the songs and the user preference, that is the average of the embeddings of all songs he/she listened to. Then, the top-N most similar songs were recommended to the user by using the cosine similarity.

\par The model proposed by \cite{Zhao2017} extends the Skip-gram model to generate the Embeddings according to the sequence of POIs visited by the user. What could be seen as a sentence in the models presented by \cite{mikolov} are treated as the sequence of check-ins that a user made. In addition to the POIs, the temporal state (weekday and weekend) in which the check-in is made was also added to the model.

\par The work done by \cite{Li2018} to generate the Embeddings is an extension of the method proposed by \cite{Barkan2016} called item2vec, which maps items for Collaborative Filtering RS into a vector representation. \cite{Li2018} extended the method by using the frequency in which the items appeared within a sequence as a weighted factor in the process of generating the Embeddings.

\par Finally, the work method proposed in \cite{Wang2018} uses the Skip-gram model as well, but it does so in two distinct ways: an user and a session-based. The user-based is used by feeding all the musics that the user has listened to as a sequence to the Skip-gram model, and the session-based splits the data into sessions, feeding them as sequence instead of the user's listening history.

\subsection*{Deep Learning Models}

\par In order to generate contextual recommendations using DL models, all of the works used information about the time in which the ratings were given, by using the Timestamp itself as an attribute or the sequence in which the ratings are in the data, using that order to extract the intrinsic behavior of the user.

\par
\begin{figure}[!ht]
    \centering
    \begin{subfigure}[b]{.55\textwidth}
        \includegraphics[width=\textwidth]{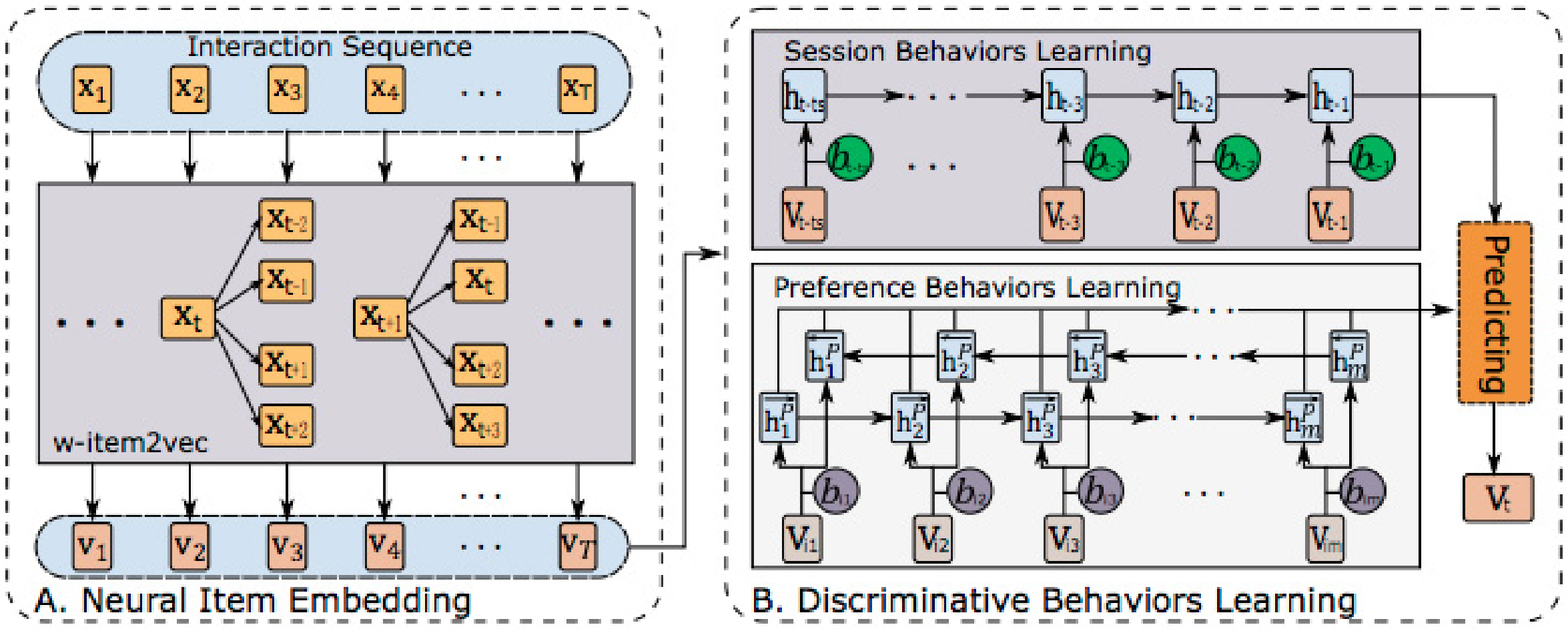}
        \caption{The embeddings model converts the sequence of input to the Deep Learning architecture \cite{Li2018}.}
        \label{fig:li}
    \end{subfigure}
    ~
    \begin{subfigure}[b]{0.35\textwidth}
        \includegraphics[width=\textwidth]{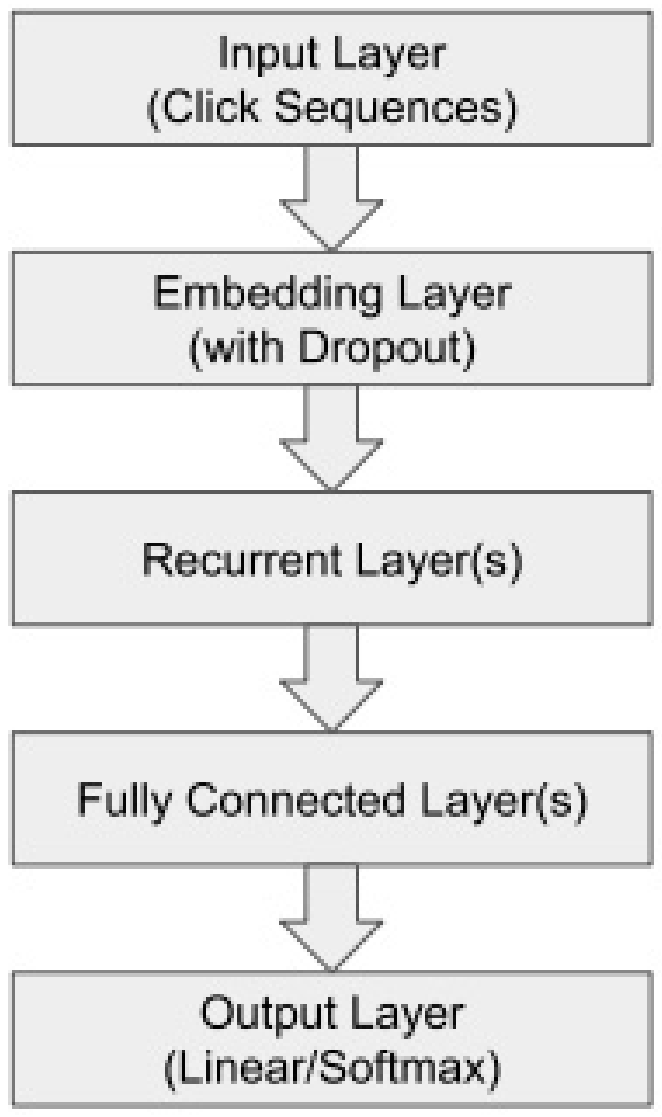}
        \caption{Architecture using Embeddings as input to the RNN \cite{Tan2016a}.}
        \label{fig:tan2}
    \end{subfigure}
    \caption{Models that used both Embeddings and Deep Learning}
\end{figure}

 The most used models to deal with sequence of data are the RNN, which were used in the following references \cite{Tan2016a, Smirnova2017, Zhang2018, Beutel2018, Li2018}. The only reference that didn't use a RNN was \cite{Jannach2017}. However, only the work done by \cite{Tan2016a} used a pure RNN to generate the recommendations while all the others did modifications in the model. \cite{Tan2016a} also used embeddings to represent the data in the input, as pointed in Figure~\ref{fig:tan2}.

\par
\begin{figure}
    \centering
    \begin{subfigure}[b]{0.49\textwidth}
        \includegraphics[width=\textwidth]{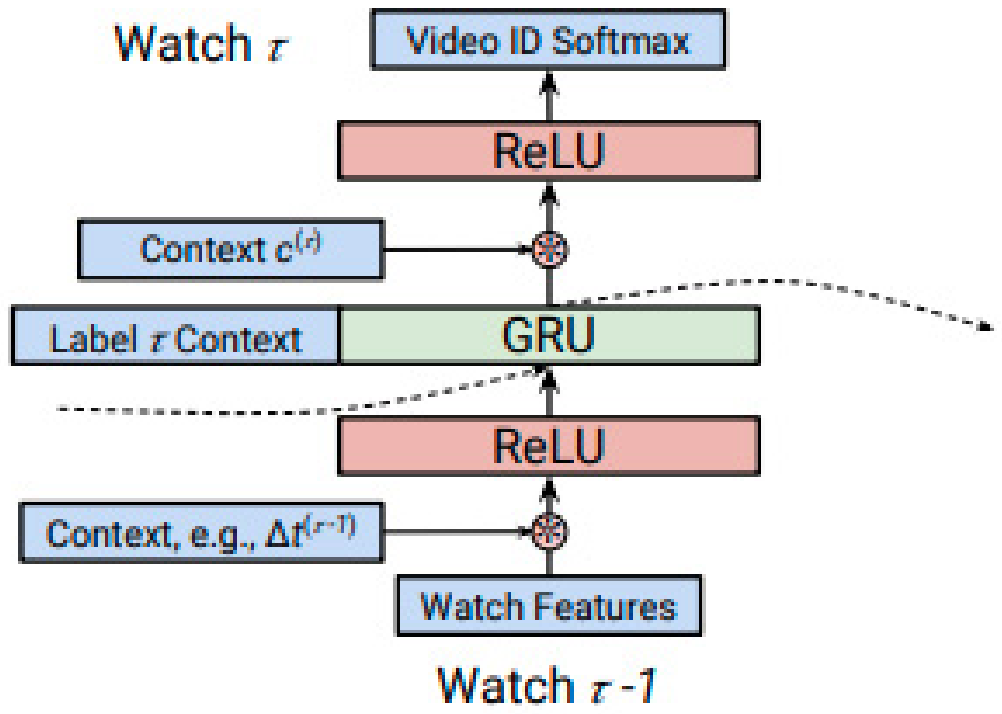}
        \caption{How the context was used in the diagram of the RNN \cite{Beutel2018}.}
        \label{fig:beutel}
    \end{subfigure}
    ~ 
    \begin{subfigure}[b]{0.49\textwidth}
        \includegraphics[width=\textwidth]{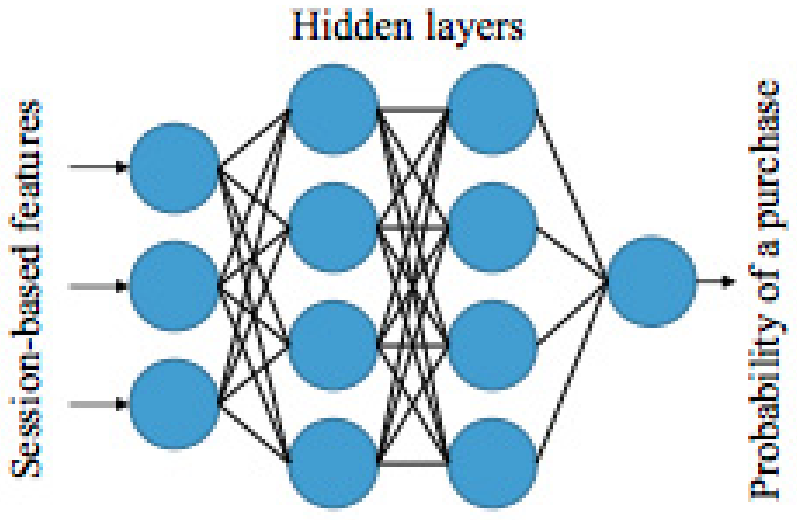}
        \caption{Feed forward Network that outputs the probability of a purchase \cite{Jannach2017}.}
        \label{fig:jannach}
    \end{subfigure}
    \begin{subfigure}[b]{.6\textwidth}
        \includegraphics[width=\textwidth]{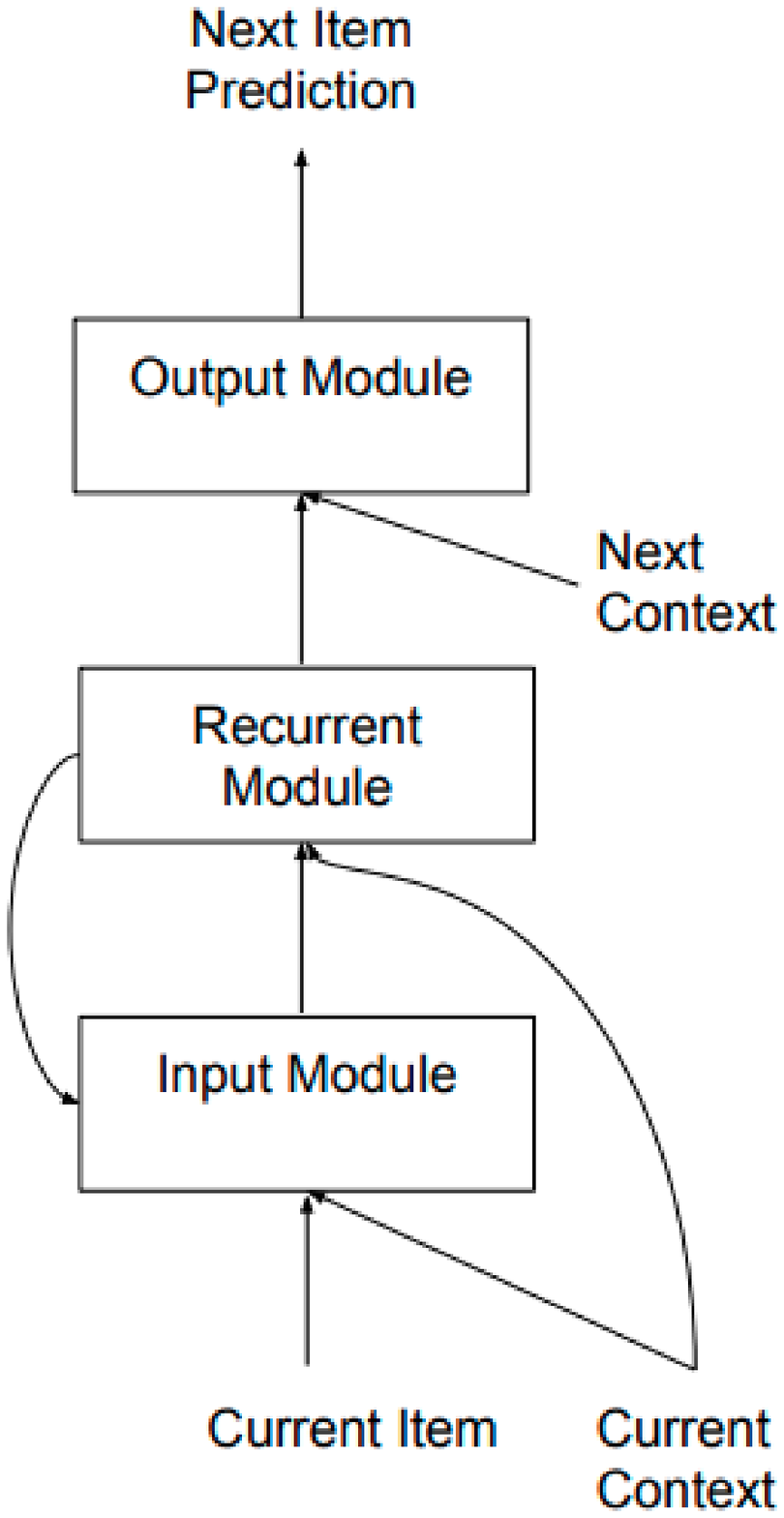}
        \caption{How the contextual sequence is modeled in the recurrent architecture \cite{Smirnova2017}.}
        \label{fig:smirnova}
    \end{subfigure}
    \caption{Models that use DL to generate the recommendations}\label{fig:dl_models}
\end{figure}

 The work proposed by \cite{Smirnova2017} changed the model of the RNN in two ways: (1) By changing the representation of the item, creating a representation that contained both contextual information and information about the item, in order to capture context-dependent item similarities, and (2) They conditioned the hidden dynamics of the RNN to work with contextual information. How the contextual information was used can be seen in Figure~\ref{fig:smirnova}. This model was used to deal with a recommendation strategy called Next-Item Prediction, in which the recommender must predict the next item that a user will most-likely enjoy, based on the past interactions of the user. 

\par The work proposed by \cite{Beutel2018} used RNNs with a technique known as \textit{Latent Cross}, which first uses different contextual information in the RNN model. The Latent Cross technique embed the contextual information before and after the hidden states, by performing an element-wise product of the context embedding with model’s hidden states. They used this model in the video domain, using contextual information about which page and device the video was played, as well as the time. Figure~\ref{fig:beutel} presents the general architecture of the model.

\begin{figure}[!ht]
    \includegraphics[width=\textwidth]{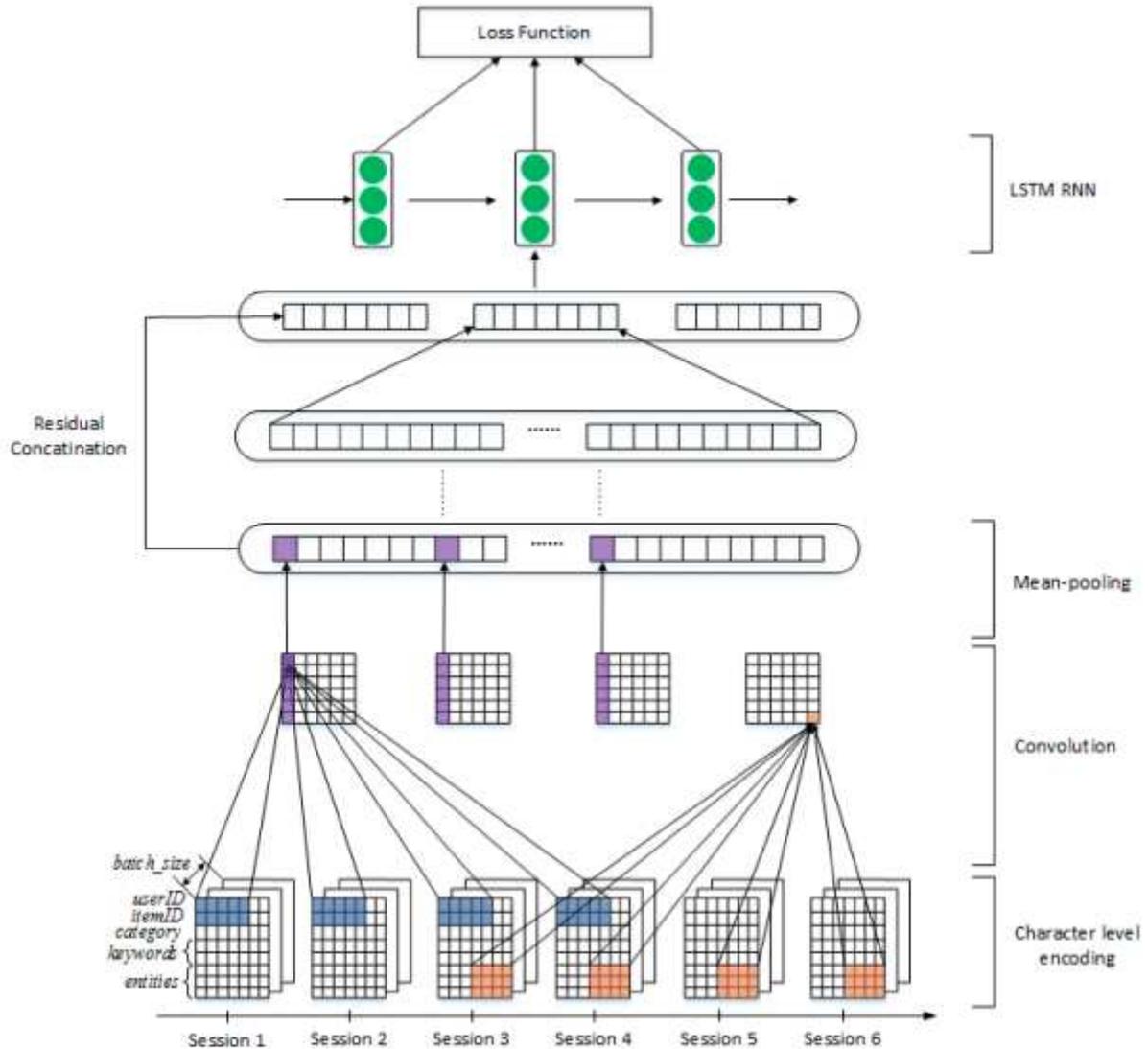}
    \caption{The architecture with a convolutional layer and an LSTM network \cite{Zhang2018}.}
    \label{fig:zhang}
\end{figure}

\par In \cite{Zhang2018}, a combination of two DL techniques (presented in Figure~\ref{fig:zhang}) were used: CNN and RNN. The model that unified both techniques to generate the recommendation was called DeepJoNN, and it is used to recommend news to users. Their model first encodes the input data in a character-level representation of a matrix. Then, a CNN is used to extract patterns of features from streams of click events within sessions. Finally, the RNNs are used to generate the recommendations from the patterns found by the CNN, predicting the top-N most valuable items to the user. In order to address the problem of exploding/vanishing gradient when learning long-term dependencies, Long Short-Term Memory (LSTM) units were used to optimize the traditional RNN.

\par In the work done by \cite{Li2018}, they have used both Embeddings and DL models to improve their CARS, as shown in Figure~\ref{fig:li}. The output of the Embeddings phase is used as the input of the DL model, which consists in two RNNs with LSTM units. The first RNN, called \textit{Preference Behaviors Learning}, is used to learn what part of the user's behavior imply its stable historical preferences. It does so by using a Bi-Contextual LSTM (B-CLSTM), inspired by the Bi-directional RNN proposed in \cite{Schuster97}, to make use of the long-term representation of the context in both forward and backward direction. The second RNN is called \textit{Session Behaviors Learning}, and its goal is to learn short-term session behavior of the user, that represents the user present's motivation. As they only want to learn the present user's preferences, they use a simple Contextual LSTM instead of a B-CLSTM.

\par The only work found that used DL models that did not use any type of RNNs is the one proposed by \cite{Jannach2017}, which uses Feed Forward Neural Networks (FFNN), as displayed in Figure~\ref{fig:jannach}. In this work, the DL model was applied as a post-processing phase to a traditional RS that take into account features of the items, in order to re-rank the results obtained by the RS. The model consider each item as an unlabeled example and then generate values for all the features given the current session, with an output as to the probability that the user will buy the item. The top-N results generated by the RS are re-ranked accordingly to the probability obtained in the output of the FFNN.

\section{ Conclusion}
\par Recommender Systems are tools to filter the information that is available to a user by using its preferences. The use of contextual information in the process of generating the recommendations can improve the algorithm because the preferences and the behavior of a user depends on the context in which he/she is. In this work, a systematic review was conducted with the goal of understanding how CARS are using techniques that accomplished state-of-the-art results in different research areas to improve its recommendations, either by using it to obtain the context or to generate the recommendations. We analyzed 16 references from different data sources, and from them we can conclude that the area of Context-Aware Recommender Systems has recently been using Deep Learning, Embeddings and even further the combination of both techniques to improve the recommendations, obtaining great success, especially in the applications that deal with recommending items to an user by analyzing the user's session.
\par From our systematic review, we can conclude that there are some trends in the models used, both in Deep Learning and Embeddings. The most used DL model are the Recurrent Neural Networks, which are designed to be effective when dealing with sequences of data. The RNN models were effective to obtain contextual information from the sequence of input data, and outperformed the baseline models in different domains, showing that those models can be used as CARS. Word2Vec was the most used model, and was modified by the almost all of the references, according to the domain. Distributed representations, generated both by Embeddings or Deep Learning models were capable of extracting contextual information from the data, improving the recommendations obtained by CARS. One of the reasons of its success is the possibility to calculate the similarity between the items more easily, as they are presented as vectors of real numbers. 

\section*{Acknowledgements}
\par The authors would like to thanks CAPES/CNPq for financial support.

\bibliographystyle{unsrt}
\bibliography{references}  

\end{document}